# Isotopic variation of parity violation in atomic ytterbium

D. Antypas[1,*], A. Fabricant[2], J.E. Stalnaker[3], K. Tsigutkin[4], V.V Flambaum[2,5] and D. Budker[1,2,6]

[1]Helmholtz-Institut Mainz, Mainz 55128, Germany
[2]Johannes Gutenberg-Universität Mainz, Mainz 55128, Germany
[3]Department of Physics and Astronomy, Oberlin College, Oberlin, Ohio 44074, USA
[4]ASML, Veldhoven, The Netherlands
[5]School of Physics, University of New South Wales, Sydney 2052, Australia
[6]Department of Physics, University of California at Berkeley, California 94720-300, USA

**The weak force is the only fundamental interaction known to violate the symmetry with respect to spatial inversion (parity). This parity violation (PV) can be used to isolate the effects of the weak interaction in atomic systems, providing a unique, low-energy test of the Standard Model [see for example reviews 1, 2, 3]. These experiments are primarily sensitive to the weak force between the valence electrons and the nucleus, mediated by the neutral $Z^0$ boson and dependent on the weak charge of the nucleus, $Q_w$. The Standard Model (SM) parameter $Q_w$ was most precisely determined in cesium (Cs) [4, 5] and has provided a stringent test of the SM at low energy. The SM also predicts a variation of $Q_w$ with the number of neutrons in the nucleus, an effect whose direct observation we are reporting here for the first time. Our studies, made on a chain of ytterbium (Yb) isotopes, provide a measurement of isotopic variation in atomic PV, confirm the predicted SM $Q_w$ scaling and offer information about an additional Z´ boson.**

The large PV observable in Yb was first predicted by DeMille [6], a prediction further supported by subsequent calculations [7,8,9] and confirmed by experiment [10,11]. The PV effect in Yb is approximately 100 times larger than that in Cs. Moreover, Yb has a chain of stable isotopes, allowing for an isotopic comparison of the effect [12]. Such a comparison has the potential to be a probe of neutron distributions in the Yb nuclei [13] and is sensitive to physics beyond the SM [14, 15]. A related measurement, in which the PV effects are compared for different hyperfine components of isotopes with non-zero nuclear spin, is expected to improve the understanding of the weak interaction within the nucleus [3,16,17,18].

The principle of our measurements is similar to that of the 1st-generation experiment [10,11]. We optically excite Yb atoms in a beam, on the $6s^2\ ^1S_0 \rightarrow 5d6s\ ^3D_1$ transition (fig. 1), in a region in which in addition to the applied optical field, static electric and static magnetic fields are applied to the atoms [19]. The directions of the magnetic and static electric field and that of the optical-field polarization define the handedness for the experimental coordinate system. As the $^1S_0$ and $^3D_1$ states are of nominally same parity, an electric-dipole (E1) transition between them is forbidden by selection rules. In the presence of the weak interaction, however, mixing of the $^1P_1$ state into $^3D_1$ results in a E1 PV amplitude for the transition. The applied dc (or quasi-static) electric field results in additional mixing of these states, allowing for a larger and controlled Stark-induced E1 amplitude [20]. The Stark-induced and PV amplitudes will interfere with appropriate choice of field geometry. Field reversals flip the handedness of the field geometry, leading to a sign reversal of the Stark-PV interference term and a change in the transition rate. This change provides an experimental observable.

We measured the PV effect in four nuclear-spin-zero isotopes ($^{170}$Yb, $^{172}$Yb, $^{174}$Yb and $^{176}$Yb, with abundances 3.1%, 21.9%, 31.8% and 12.7%, respectively). In each of these isotopes we excited the $m=0 \rightarrow m'=0$ component of the $^1S_0 \rightarrow\ ^3D_1$ transition ($m$ denotes the magnetic sublevel of a state), which is spectrally separated from the other two components $0 \rightarrow \pm 1$ with application of a magnetic field (see Methods section). This separation is done because the Stark-PV interference averages to zero over the three transition components. The transition is driven with linearly polarized light at 408 nm, propagating along $x$ (see fig. 2) with components $\mathcal{E}_y=\mathcal{E}\sin\theta$, $\mathcal{E}_z=\mathcal{E}\cos\theta$, where $\vec{\mathcal{E}}$ is the optical field and $\theta$ is the angle between $\vec{\mathcal{E}}$ and $+z$. The required electric field $\vec{E}$ is applied along $\pm x$. This field consists of an oscillating component $E_0\cos\omega t$ ($E_0$ in the range 0.8-1.6 kV/cm, $\omega/2\pi =19.9$ Hz), and a dc component $E_{dc} \approx 6$ V/cm that helps control systematic errors in detection of the Stark-PV interference. The magnetic field $\vec{B}$ required to observe this interference is along ±z. With this field geometry the Stark-PV interference term is proportional to a parity-odd and time-reversal-even pseudo-scalar rotational invariant given by [19,21]:

$$(\vec{\mathcal{E}} \cdot \vec{B})([\vec{E} \times \vec{\mathcal{E}}] \cdot \vec{B}). \tag{1}$$

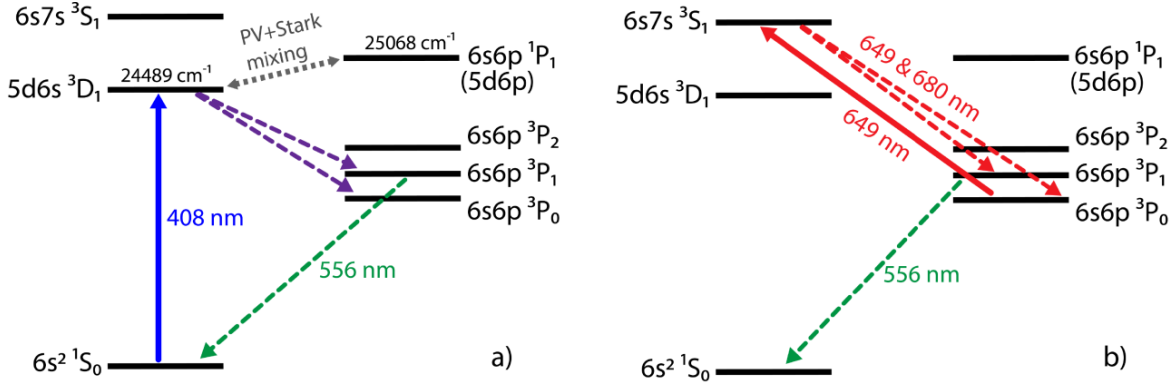

**Fig. 1. Energy-level diagram of Yb with relevant optical transitions.** a) The weak-force-induced mixing between the $^3D_1$ and $^1P_1$ states introduces an admixture of the odd-parity $^1P_1$ state into the nominally even-parity $^3D_1$ state, allowing the otherwise-forbidden $^1S_0 \rightarrow {}^3D_1$ excitation at 408 nm to proceed with an E1 amplitude. The mixing is enhanced due to the proximity of the two energy levels and due to significant admixture of the 5d6p configuration in the $^1P_1$ level [6]. Additional (Stark) mixing between the $^3D_1$ and $^1P_1$ states is introduced with an electric field. About 2/3 of atoms undergoing the 408 nm transition relax to the $^3P_0$ metastable level. b) Transitions are detected by monitoring the population in the $^3P_0$ via fluorescence resulting from further excitation to the $^3S_1$ state.

The resulting excitation rate contains terms oscillating at $\omega$ and $2\omega$ (see Methods section). Phase-sensitive detection at these frequencies yields the ratio of harmonics:

$$r_0 = \frac{4E_{\text{dc}}}{E_0} + \frac{4\zeta}{\beta E_0} \cot\theta, \tag{2}$$

where $\beta$ is the vector polarizability of the transition and $\zeta$ is the imaginary PV-induced E1 transition moment [11], a quantity approximately proportional to $Q_w$. We measure $\zeta/\beta$ for given $E_0$ by extracting the part of the ratio $r_0$ that modulates with the polarization angle ($\theta = \pm\pi/4$), and averaging that part over opposite $B$ and $E_{\text{dc}}$. For the typical field $E_0 = 1.2$ kV/cm, this modulation in $r_0$ is of amplitude $4\zeta/\beta E_0 \approx 8 \cdot 10^{-5}$. In addition to the $E$ and $\theta$ reversals that discriminate the PV signal from parity-conserving background, reversing $B$ and $E_{dc}$ provides us with important handles to characterize and minimize systematic contributions not explicitly shown in equation (2).

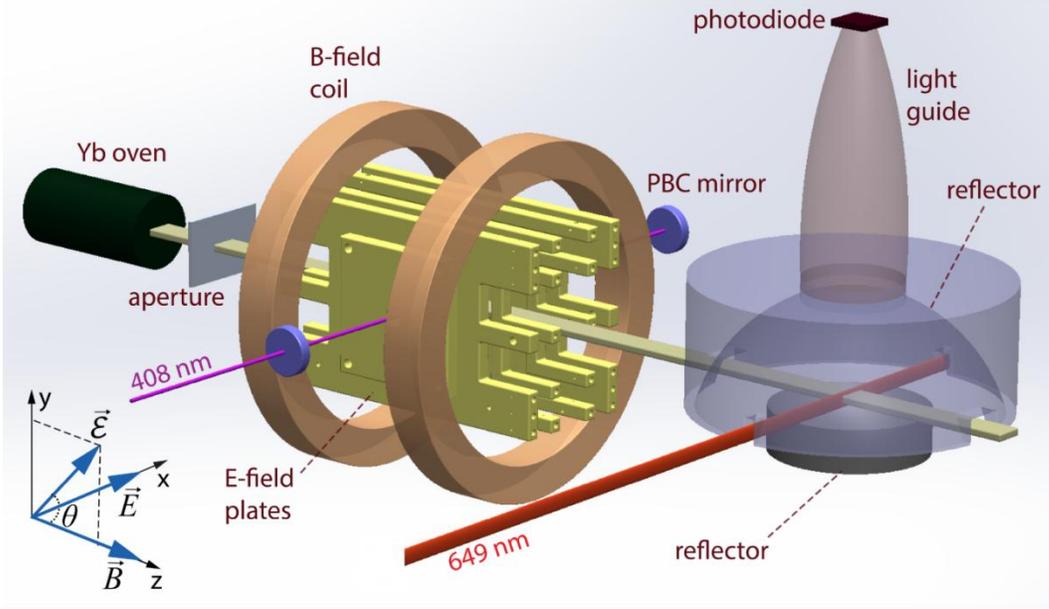

**Fig. 2. Schematic of the Yb atomic-beam apparatus**. An atomic beam is produced under vacuum with an oven heated to ≈600 °C. The 1.5 cm wide and 0.6 mm high interaction region, where the $^1S_0 \rightarrow {}^3D_1$ excitations occur, is defined by the intersection of the atomic beam with the resonant light beam of the power-build-up cavity (PBC), whose finesse is ≈550. A system of gold-coated electrodes is used to create an electric field in the interaction region. This system consists of two ≈10x10 cm$^2$ parallel plates, spaced by 5.5045(20) cm, and a set of surrounding electrodes used to increase field uniformity as well as to apply additional field components for systematics studies. A pair of coils is employed to create a 93 G magnetic field, large enough to fully resolve the three components of the 408 nm resonance, whose Doppler-broadened width is ≈18 MHz. Most of the atoms excited to $^3D_1$ decay to the $^3P_0$ metastable state, whose population is probed downstream from the interaction region through further excitation to the $^3S_1$ state with ≈120 μW of laser light at 649 nm, and detection of the induced fluorescence at 556, 649 and 680 nm (see fig. 1b). Fluorescence is collected with a light collector and measured with a large-area photodiode. This detection scheme is far more efficient than direct detection of fluorescence from the interaction region. The latter is not done in this experiment.

Our atomic-beam apparatus (shown in fig. 2) is described in detail in [22]. Yb atoms effusing from an oven reach the interaction region and are exposed to 408 nm light circulating in a power-build-up cavity (PBC). This cavity is used to enhance the $^1S_0 \to {}^3D_1$ excitation rate and suppress unwanted interference between the Stark amplitude and that of a magnetic-dipole (M1) transition [23]. The 408 nm light coupled to the cavity is the output of a frequency-doubled Ti:Sapphire laser. The resonance frequency of the PBC is locked to the Ti:Sapphire frequency, which is in turn stabilized to the peak of the $^1S_0 \to {}^3D_1$ resonance (see Methods section). It is necessary to actively stabilize the circulating power ($\approx$55 W) in the PBC to avoid noise in the $^1S_0 \to {}^3D_1$ excitation rate. A set of electric-field plates and magnetic-field coils are used to apply the required fields to the atoms. Additional coils are employed to null the earth field as well as to apply field components to study and control systematic contributions to the PV measurements. The polarization of the 408 nm light is controlled with a half-wave plate mounted on a motorized rotation stage, and is monitored with a polarimeter measuring light transmitted through the PBC. The $^1S_0 \to {}^3D_1$ transition rate is measured downstream from the interaction region using a shelving technique, described in detail in [11,22]. Phase-sensitive detection of the $\omega$ and $2\omega$ harmonics present in this rate is done with lock-in amplifiers.

We acquired PV data totaling 34 days of measurement time during a 2.5 month period, for a total of 260 hr of actual integration time. At regular intervals, we measured PV-mimicking systematics, to confirm that these are always sufficiently small. In each of our 900 PV runs (each requiring $\approx$1/2 hr), the ratio of equation (2) was measured 200 times under all polarity combinations of $E$, $\theta$, $B$ and $E_{dc}$, which were reversed at a rate of 19.9, 0.12, 0.06, and 0.03 Hz respectively. We applied an electric field of amplitude $E_0$=0.8 or 1.2 kV/cm (1.2 or 1.6 kV/cm with $^{170}$Yb) and alternated between isotopes, to minimize the impact of potential drifts. A set of $\zeta/\beta$ values was obtained for each run, with a statistical error that ranges from 5% for $^{174}$Yb to 16% for $^{170}$Yb.

The signal-to-noise ratio (SNR) in measuring the PV effect is $\approx 0.55\sqrt{t(s)}$ ($t$ is the integration time) for the highest-abundance isotope ($^{174}$Yb), about 18 times better than that of the 1$^{st}$-generation experiment [10,11]. The improvement is mainly due to lower frequency noise of the light circulating in the PBC in the present apparatus, and reduced sensitivity to this noise with the current measurement scheme, which involves acquisition at the peak of the $^1S_0 \rightarrow {}^3D_1$ resonance profile (see Methods section), rather than the lineshape fitting done in the previous experiment. The attained sensitivity is roughly consistent with shot-noise-limited detection of the estimated 408 nm transition rate in the atomic beam. The observed SNR scaling with isotopic abundance shows that the detection of the Stark-PV interference approaches the shot noise limit.

A number of corrections must be applied to the raw data. The largest of these are due to slight saturation of the $^1S_0 \rightarrow {}^3D_1$ transition rate (0.5-1.2%), and due to the transit time of atoms from the interaction to the detection region (0.3%) [11]. A smaller correction (0.05-0.15%) accounts for slight deviations of $\theta$ from $\pm\pi/4$.

Systematic errors arise from uncertainties in calibrations applied to the data, or uncertainties in determining false-PV signals. Signals mimicking the true PV effect come from stray (non-reversing) fields and field misalignments. Through careful control of these field imperfections, we keep PV-mimicking contributions and the associated uncertainties well below 0.1% of $\zeta/\beta$. The dominant calibration error comes from an unexpected drift in electronics processing the experimental signal prior to lock-in detection (0.22%). The second-largest error is due to imperfections in measuring polarization (0.1%). In table I, we list leading systematics in determining $\zeta/\beta$.

Table I: Main systematic errors in $\zeta/\beta$ measurements.

| Contribution | Error (%) |
|---|---|
| Harmonics-ratio calibration | 0.22 |
| Polarization angle | 0.1 |
| High-voltage measurements | 0.06 |
| Transition saturation correction | 0.05* |
| Field-plate spacing | 0.04 |
| Stray fields & field misalignments | 0.02 |
| Photodetector response calibration | 0.02 |
| **Total** | **0.26** |

*0.09 for $^{170}$Yb. The error is larger because data for this low-abundance isotope were taken at a higher electric field.

Three kinds of experiments were done to investigate potentially unaccounted-for systematic effects and ensure measurement consistency under various conditions. First, data were acquired with enhanced field imperfections, which were 2 to 3 orders of magnitude greater than those during actual PV-data acquisition, as well as with either the left or right half of the atomic beam blocked, to check for the effects of potential field gradients. The results of these experiments (with uncertainty ≈3% in the PV-effect) did not reveal the presence of unaccounted for PV-mimicking effects. Second, a set of PV-runs was done with changing apparatus components, such as with different field plates and high-voltage amplifiers, and without the PBC cavity. These experiments, of uncertainty ≈3%, 3% and 25%, respectively, were consistent with the final result for PV-effect. Finally, several consistency checks were carried out, including measurements with the laser frequency tuned to the side of the 0→0 408 nm transition profile and on the 0 → ±1 components of the transition. Fitting the transition lineshape to determine $\zeta/\beta$ (as in the 1[st]-generation experiment) was done to check for systematic effects associated with the transition profile. Data were also acquired on the $F=1/2 \rightarrow F´=1/2$ component of the $^{171}$Yb 408 nm transition ($F$ denotes the total angular momentum of a state), where there is no PV-observable [24]. All these consistency checks, to within their statistical uncertainty (1% to 5% depending on the experimental test), did not reveal unaccounted-for systematic effects.

The result for $\zeta/\beta$ in each of the isotopes is (in mV/cm): $^{170}$Yb: -22.81(22), $^{172}$Yb: -23.24(10), $^{174}$Yb: -23.89(11), $^{176}$Yb: -24.12(10). The error in parenthesis is statistical; the systematic uncertainty is 0.06 mV/cm in all isotopes. The reduced $\chi^2$ values associated with these data sets are 1.09, 0.92, 0.99, and 1.02, respectively.

A plot of $-\zeta/\beta$ values vs. neutron number is shown in fig. 3a. We observe a clear isotopic dependence of the PV effect that can be compared with the SM model prediction for $Q_W$, which to lowest order is given by $Q_W = -N + Z(1 - 4\sin^2\theta_W)$ [1], where $N$ and $Z$ are the neutron and proton numbers, respectively, and $\theta_W$ is the Weinberg angle. A higher-accuracy expression for $Q_W$ (accurate at the 0.1% level) is obtained with inclusion of radiative corrections [25]:

$$Q_W = -0.989 \cdot N + 0.071 \cdot Z. \tag{3}$$

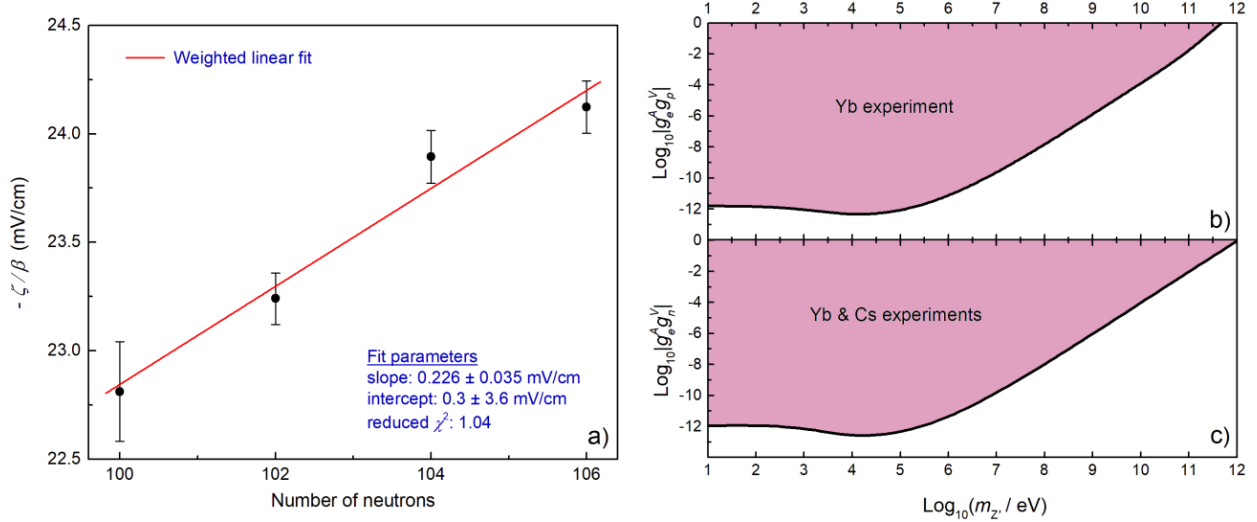

**Fig. 3. a) Isotopic variation of $\zeta/\beta$**. Error bars represent the statistical and systematic error combined in quadrature. Uncertainties in the parameters of the weighted fit to the data refer to the 1σ errors. The weight assigned to each data point is the inverse square of the corresponding error bar shown in the plot. The intercept of the fitted line that depends on the contribution of the protons is small; it is consistent with the expected ≈−1.2 mV/cm proton contribution to −$\zeta/\beta$, which provides additional confirmation that the effect measured in this experiment is indeed due to neutrons (i.e. due to the weak charge). This expected proton contribution is estimated as follows: the weak charge carried by protons is $Q_p$≈5 (see text) which, for the mean neutron number $N$=103 in these measurements, corresponds to a fraction of the total weak charge of equation (3) of $Q_p/Q_W$ ≈ *−0.0515* or to a contribution to −$\zeta/\beta$ of $(-\zeta/\beta)_p=(Q_p/Q_W)\cdot(-\zeta/\beta)_{N=103}$ ≈−1.2 mV/cm. **b) Bounds on light Z´-mediated PV electron-proton interactions.** The black line represents the 1σ limit on the particular coupling, shown for a large range of the boson mass $m_{Z'}$. The colored region in the plot corresponds to the parameter space excluded by the Yb experiment. The low-mass ($m_{Z'}$ <100 eV/c²) limit for the coupling $|g_e^A g_p^V| = 1.6\cdot 10^{-12}$, and the corresponding large-mass asymptotic limit ($m_{Z'}$ >100 MeV/c²) is $|g_e^A g_p^V|/m_{Z'}^2 = 1.3\cdot 10^{-6}$ (GeV)$^{-2}$. **c) Bounds on light Z´-mediated PV electron-neutron interactions.** This result comes from combining existing limits on the effective electron-nucleon coupling, derived from the Cs PV experiment [4], with the Yb experiment limits shown in b). The low-mass limit for the interaction is $|g_e^A g_n^V| = 1.2\cdot 10^{-12}$, and the corresponding large-mass asymptotic limit is $|g_e^A g_n^V|/m_{Z'}^2 = 9.3\cdot 10^{-7}$ (GeV)$^{-2}$.

The proton contribution to $Q_W$ is $Q_p \approx 5$. The predicted fractional change in $Q_W$ per neutron is $(1/Q_W)\cdot(dQ_W/dN) \approx 1\%$ around $N=103$ (mean neutron number for isotopes measured). Using the parameters of a weighted linear fit to the data of fig. 3a, we obtain the observed fractional variation of $-\zeta/\beta$: slope/$(-\zeta/\beta)_{N=103}$=0.96(15)% per neutron. This observed variation agrees well with the SM expectation, offering a confirmation of the predicted $Q_W$ dependence. The expected neutron-skin-related contribution to the $Q_W$ variation between the two extreme isotopes (N=100 and 106) measured is ≈0.1% of $Q_W$ [14], well below our current sensitivity in measuring this variation.

Our measurements allow for exploration of PV electron-nucleon interactions mediated by a light vector boson Z´. Such couplings result in a beyond SM contribution to $Q_W$. Several searches for vector bosons of mass $m_{Z'} > 100$ keV/c$^2$ have been carried out, as well as searches for interactions of SM matter with dark-matter bosons and dark-energy fields (see, for example, review [3] and references therein). Upper bounds on exchange of a light Z´ (of lowest mass considered ≈10$^{-20}$ eV/c$^2$) exist, and come from torsion-pendulum [26,27] and atomic-magnetometry [28] experiments, as well as from atomic calculations [15] that utilize results of previous atomic PV experiments. The present isotopic-comparison data can be analyzed to provide individual constraints on electron-proton and electron-neutron interactions. The extracted proton contribution to $\zeta/\beta$ (fig. 3a) is used to set limits on the former (see Methods). This result is then combined with existing bounds on the sum of electron-proton and electron-neutron couplings to constraint electron-neutron couplings. These constraints are shown in fig. 3b and fig. 3c.

The ratio $\zeta/\beta$ was reported in [10,11] with a sign opposite to that reported here. Newly obtained data were analyzed using the data-processing scripts employed in [10,11] that allowed us to trace a sign-related error in the previous data-analysis procedure. A number of further tests done with the current apparatus (see Methods), are in support of a correct sign determination for $\zeta/\beta$ in the present work. The magnitude of $\zeta/\beta$ for $^{174}$Yb was previously reported as 39(4)$_{stat}$(5)$_{syst}$ mV/cm, which is a ≈2.4σ difference from our much higher-accuracy result. This discrepancy points to a possible underestimation of systematic uncertainties. Note that the 1$^{st}$-generation apparatus achieved an SNR in the PV-effect measurements of ≈18 times lower than that in the current experiment, making high-accuracy studies of systematic effects more difficult.

The single-isotope accuracy in this experiment is ≈0.5% for three of the Yb isotopes measured, approaching that of the most precise measurements to date in Cs (0.35%). The current accuracy level allows a definitive observation of the isotopic variation of the weak force between the nucleus and valence electrons. Furthermore, the measured variation agrees well with the prediction by the electroweak theory regarding the weak-charge scaling with the number of neutrons. Analysis of the obtained PV-data is used to constrain separately electron-proton and electron-neutron interactions that are mediated by a light Z´ boson. Our results also serve as an important sensitivity benchmark for our 2nd-generation atomic-beam apparatus. Further sensitivity upgrades are currently underway. These include an increase in the atomic beam flux, laser cooling of the atomic beam, as well as optical pumping of the non-zero nuclear spin isotopes. These upgrades, combined with tighter control of systematics, will enhance the measurement sensitivity to a level that allows high-precision isotopic comparison to probe the neutron distributions in the Yb nucleus and potentially physics beyond the Standard Model. As shown in [14], the latter searches should not be hindered by uncertainties in the knowledge of Yb nucleus neutron skin. In addition, our planned investigation of spin-dependent parity violation through a comparison of the effect among the different hyperfine levels of the forbidden transition in the non-zero-spin isotopes will be a platform to study the hadronic weak interaction.

**Acknowledgments**

We thank M. Safronova, M. Kozlov, S. Porsev, M. Zolotorev, A. Viatkina, L. Bougas and N. Leefer for useful discussions. AF is supported by the Carl Zeiss Graduate Fellowship.

**References**

1. Ginges, J.S.M., and Flambaum, V.V. Violations of fundamental symmetries in atoms and tests of unification theories of elementary particles. Phys. Rep. **397**, 63 (2004).
2. Roberts, B.M., Dzuba, V.A., and Flambaum, V.V. Parity and time-reversal violation in atomic systems. Annu. Rev. Nucl. Part. Sci. **65**, 63 (2015).
3. Safronova, M.S. et al. Search for new physics with atoms and molecules. Rev. Mod. Phys. **90**, 025008 (2018).
4. Wood, C.S. et al. Measurement of parity nonconservation and an anapole moment in cesium. Science **275**, 1759 (1997).
5. Dzuba, V. A., Berengut, J. C., Flambaum, V. V., and Roberts, B. Revisiting parity nonconservation in Cesium. Phys. Rev. Lett. **109**, 203003 (2012).
6. DeMille, D. Parity nonconservation in the $6s^2\ ^1S_0 \rightarrow 5d6s\ ^3D_1$ transition in atomic ytterbium. Phys. Rev. Lett. **74**, 4165 (1995).
7. Porsev, S.G., Rakhlina, Yu. G., and Kozlov, M.G. Parity violation in atomic ytterbium. JETP Lett. **61**, 459 (1995).
8. Das, B. P. Computation of correlation effects on the parity-nonconserving electric-dipole transition in atomic ytterbium. Phys. Rev. A **56**, 1635 (1997).
9. Dzuba, V. A., and Flambaum, V. V. Calculation of parity nonconservation in neutral ytterbium. Phys. Rev. A **83**, 042514 (2011).
10. Tsigutkin, K. et al. Observation of a large atomic parity violation effect in ytterbium. Phys. Rev. Lett. **103**, 071601 (2009).
11. Tsigutkin, K. et al. Parity violation in atomic ytterbium: Experimental sensitivity and systematics. Phys. Rev. A **81**, 032114 (2010).
12. Dzuba, V. A., Flambaum, V.V., and Khriplovich, I. B. Enhancement of P- and T-nonconserving effects in rare-earth atoms. Z. Phys. D **1**, 243 (1986).
13. Fortson, E.N., Pang, Y., and Wilets, L. Nuclear-structure effects in atomic parity nonconservation. Phys. Rev. Lett. **66**, 677 (1991).
14. Brown, B. A., Derevianko, A., and Flambaum, V.V. Calculations of the neutron skin and its effect in atomic parity violation. Phys. Rev. C **79**, 035501 (2009).
15. Dzuba, V.A., Flambaum, V. V., and Stadnik, Y. V. Probing low-mass vector bosons with parity nonconservation and nuclear anapole moment measurements in atoms and molecules. Phys. Rev. Lett. **119**, 223201 (2017).
16. Flambaum, V.V. and Khriplovich, I.B. P-odd nuclear forces: a source of parity violation in atoms. ZhETF **79**, 1656 (1980). JETP **52**, 835 (1980).
17. Flambaum, V.V., Khriplovich, I.B., and Sushkov, O.P. Nuclear anapole moments. Phys. Lett. B. **146**, 367 (1984).
18. Haxton, W.C., and Wieman, C.E. Atomic parity nonconservation and nuclear anapole moments. Annual Rev. Nucl. Part. Sci. **51**, 261 (2001).

19. Bouchiat M.A., and Pottier L. Optical experiments and weak interactions. Science **234**, 1203 (1986).
20. Bouchiat, M.A., and Bouchiat, C. Parity violation induced by weak neutral currents in atomic physics. Part II. J. Phys. France **36**, 493 (1975).
21. Drell, P.S., and Commins, E.D. Phys. Rev. Lett. **53**, 968 (1984).
22. Antypas, D., Fabricant, A., Bougas, L., Tsigutkin, K., and Budker, D. Towards improved measurements of parity violation in atomic ytterbium. Hyperfine Interact. **238**, 21 (2017).
23. Bouchiat, M. A., Coblentz, A., Guéna, J., and Pottier, L. Can imperfect light polarization mimic parity violation in Stark experiments on forbidden M1 transitions? J. Phys. France **42**, 985 (1981).
24. Tsigutkin, K., Stalnaker, J.E., Budker, D., Freedman, S.J., and Yashchuk, V.V. Towards measuring nuclear-spin-dependent and isotopic-chain atomic parity violation in ytterbium. From Parity Violation to Hadronic Structure and more..., Ed. by K. de Jager *et al*, Springer (2007).
25. Tanabashi, M. et al. (Particle Data Group), Phys. Rev. **D** 98, 030001 (2018).
26. Heckel, B. R. et al. New CP-violation and preferred-frame tests with polarized electrons. Phys. Rev. Lett. **97**, 021603 (2006).
27. Heckel, B. R. et al. Preferred-frame and CP-violation tests with polarized electrons. Phys. Rev. **D**. 78, 092006 (2008).
28. Vasilakis, G., Brown, J. M., Kornack, T. W., and Romalis, M. V. Limits on New Long Range Nuclear Spin-Dependent Forces Set with a K-$^3$H comagnetometer. Phys. Rev. Lett.**103**, 261801 (2009).

**Author contributions**

D. A. built the apparatus, took and analyzed data, and wrote the manuscript. A. F. contributed to the apparatus construction, took data and edited the manuscript. J. S. participated in studies of systematic errors, contributed to data analysis and edited the manuscript. K.T. participated in studies of systematics and data analysis. V.F. led the analysis of data to extract limits on Z´ boson-mediated interactions. D. B. supervised the project and edited the manuscript.

**Competing interests**

The authors declare no competing interests.

**Correspondence and requests of materials** should be addressed to D.A.

**Methods**

**Harmonics ratio of equation (2).** The expressions for the Stark- and PV-induced amplitudes in the $^1S_0$ → $^3D_1$ transition are derived in [11]:

$$A_m^{Stark} = i\beta(-1)^{m'}(\vec{E} \times \vec{\mathcal{E}})_{-m'}, \qquad (4)$$

$$A_{m'}^{PV} = i\zeta(-1)^{m'}\vec{\mathcal{E}}_{-m'}, \quad (5)$$

where $m'$ refers to the magnetic sublevel of the $^3D_1$ state and $V_q$ represents the $q$-component of the vector $V$ in the spherical basis. Given the geometry of fields in the experiment, we obtain the following transition rate $R_0$ for the 0→0 component of the transition ($m'=0$):

$$R_0 \propto \left|A_0^{Stark} + A_0^{PV}\right|^2 \cong 2\mathcal{E}^2 E_0^2 \beta^2 \sin^2\theta + 4\mathcal{E}^2 E_{dc}^2 \beta^2 \sin^2\theta + 8\mathcal{E}^2 E_{dc}\beta\zeta \cos\theta \sin\theta +$$
$$\underbrace{(8\mathcal{E}^2 E_0 \beta\zeta \cos\theta \sin\theta + 8\mathcal{E}^2\beta^2 E_0 E_{dc}\sin^2\theta)}_{R_0^{[1]}}\cos\omega t + \underbrace{2\mathcal{E}^2\beta^2 E_0^2 \sin^2\theta}_{R_0^{[2]}}\cos 2\omega t. \quad (6)$$

Only terms linear in $\zeta$ are retained in (6). The ratio of the 1st- to 2nd-harmonic amplitudes in the transition rate is:

$$r_0 \equiv \frac{R_0^{[1]}}{R_0^{[2]}} = \frac{4E_{dc}}{E_0} + \frac{4\zeta}{\beta E_0}\cot\theta. \quad (7)$$

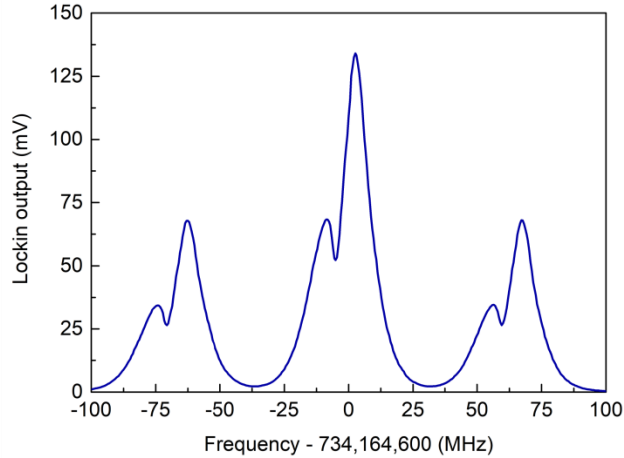

**Fig. 4. 408 nm transition spectrum**. The figure shows a profile of the $^1S_0 \to {}^3D_1$ transition, observed by scanning the frequency of the 408 nm light and recording the amplitude of the 2nd-harmonic in the excitation rate (Stark-induced rate). In the presence of the 93 G magnetic field, the magnetic sublevels of the $^3D_1$ state are shifted in energy by $m'g_J\mu_B B$, where $g_J=1/2$ is the Lande g-factor of the $^3D_1$ level and $\mu_B$ is the Bohr magneton. These energy shifts result in fully-resolved transition components. The observed lineshape asymmetry arises in the presence of the intense standing-wave field of the PBC, due to the off-resonant ac-Stark shift [29,30], and it can be suppressed if needed, using methods reported in [31].

**408 nm transition spectrum.** To observe the Stark-PV interference, the three components of the $^1S_0 \to {}^3D_1$ transition are separated with the application of a magnetic field. The resulting atomic resonance spectrum is shown in fig. 4. The PV measurements are done in the 0→0 component of the transition. To acquire data, the frequency of the laser system producing the 408 nm light is stabilized to the peak of the resonance profile.

**Apparatus improvements** The 2$^{nd}$ generation atomic beam apparatus achieved a ≈18 greater SNR in measuring the PV-effect ($0.55\sqrt{t(s)}$ for $^{174}$Yb) compared to the previous experiment (SNR reported to be $0.03\sqrt{t(s)}$ for $^{174}$Yb [11]). The improvement is primarily due to the different data acquisition scheme employed for the new measurements. Previously, the 408 nm laser frequency was swept through the complete spectral profile of the atomic resonance (all three transition components $m=0 \to m'=0,\pm 1$ of the $^1S_0 \to {}^3D_1$ were recorded), and lineshape fitting was performed to extract the parameter $\zeta/\beta$. In the new apparatus, data were taken with the 408 nm laser frequency-stabilized to the peak of the $m=0 \to m'=0$ transition, resulting in greater effective transition rate and reduced sensitivity to frequency noise of the light circulating in the PBC. When acquiring data in the new apparatus using the swept frequency scheme employed in the 1$^{st}$ generation experiment, the corresponding SNR was $0.06\sqrt{t(s)}$. Additional SNR improvement comes from the larger intracavity 408 nm light power (≈55 W) in the new apparatus, approximately seven times larger than that of the previous work.

Better uniformity of the applied electric and magnetic fields in the interaction region was achieved in the current experiment, compared to [10,11]. The quoted systematic uncertainty in the old experiment was dominated by the uncertainty in the electric field (~8%). That field was produced by arrays of wire electrodes, making precise construction challenging and reliable modeling of the applied field difficult. The electric field-plate assembly shown in fig. 2 consists of solid electrodes, and has larger dimensions (for instance a ≈5.5 cm spacing between the main conductors) compared to the old field-plate system (≈ 2 cm spacing between the main conductor surfaces). This allows for more precise construction of the plate system and more accurate modeling of the electric field (sub-0.1% uncertainty in the field). In this system, the 24 dielectric spacers placed between the 10 different conductors of the assembly are far from the interaction region, in positions that preclude direct line of sight between the spacers and the interaction region. This is a major difference of this field-plate system compared to the old one, ensuring reduced sensitivity of the measurement to stray electric fields due to charging of the spacer surfaces. We regularly measure stray fields in the new apparatus of less than 50 mV/cm, a level of control several times better than that of [10,11]. The primary magnetic field coil assembly in the new apparatus is significantly larger in size than that used in the 2009 experiment, resulting in better field uniformity in the interaction region. Additional sets of coils are employed in the current setup to characterize and control false-PV systematics, as opposed to the previous experiment in which, due to the lower experimental sensitivity, such studies were not carried out.

**Bounds on Z´-mediated PV-couplings** We consider the PV-effects arising from the exchange of a vector boson Z´ between the electrons and nucleons, in the presence of the following interaction:

$$\mathcal{L}_{int} = Z'_\mu \sum_{f=e,p,n} \bar{f}\gamma^\mu (g_f^V + \gamma_5 g_f^A) f. \tag{8}$$

With such an interaction, the PV-induced transition moment is given by [15]:

$$\zeta = E_\infty \left(\frac{-Q_W}{N}\right) + E(m_{Z'}) \frac{2\sqrt{2} A g_e^A g_N^V}{N G_F m_{Z'}^2}, \tag{9}$$

where $E(m_{Z'})$ is the calculated PV-amplitude induced by the interaction, $E_\infty$ is the value of this amplitude for an interaction mediated by SM Z boson which has mass $m_Z$, very large on the atomic scale. $G_F = 1.166 \cdot 10^{-5}$ GeV$^{-2}$ is the Fermi constant, and $g_e^A g_N^V$ is an effective axial electron-vector nucleon coupling, defined as:

$$g_e^A g_N^V \equiv (Z/A) \cdot g_e^A g_p^V + (N/A) \cdot g_e^A g_n^V, \tag{10}$$

where $A=N+Z$. The coupling $g_e^A g_p^V$ arises from electron-proton and $g_e^A g_n^V$ from electron-neutron interactions. Such interactions effectively modify the SM weak charge: $Q_W = Q_W^{SM} + Q_W^{new}$, with:

$$Q_W^{new} = -\frac{E(m_{Z'})}{E_\infty} \frac{2\sqrt{2} A g_e^A g_N^V}{G_F m_{Z'}^2}. \tag{11}$$

The amplitudes $E(m_{Z'})$ are calculated in [15] for a range of masses (10 eV/c$^2$-1 GeV/c$^2$) in several systems of interest for atomic PV studies, using the same approaches as those in standard PV-effect calculations, but allowing for large interaction range for the vector boson. Note that for $m_{Z'}$<100 eV, $E(m_{Z'})$ practically doesn't depend on $m_{Z'}$ and for $m_{Z'}$>100 MeV, $E(m_{Z'})$ is proportional to $(m_{Z'})^{-2}$. Therefore the results are defined for any $m_{Z'}$.

We use the observed proton contribution to $\zeta/\beta$ (fig. 3a) to constrain $g_e^A g_p^V$. From the y-intercept of the plot of fig. 3a), we obtain $(-\zeta/\beta)_p$=(0.3±3.6) mV/cm, from which the measured weak charge of the protons of the Yb nucleus is computed: $Q_p = Q_W^{SM} \cdot (-\zeta/\beta)_p / (-\zeta/\beta) = 1.24 \pm 14.83$. (We have used the value of $-\zeta/\beta$=23.52 mV/cm that corresponds to N=103, and calculated with equation (3) a weak charge $Q_W^{SM} = -96.88$. The effects of the neutron skin and its variation across the chain of measured isotopes were not considered here, since these only shift the value of $(-\zeta/\beta)_p$ by an estimated −0.4 mV/cm, much smaller than the 3.6 mV/cm error of its present determination.) The corresponding beyond SM contribution is $Q_p^{new} = Q_p - Q_p^{SM} = -3.75 \pm 14.83$. Equation (10) yields:

$$g_e^A g_p^V = -Q_p^{new} \frac{E_\infty}{E(m_{Z'})} \frac{G_F m_{Z'}^2}{2\sqrt{2} Z}. \tag{12}$$

The upper bounds for $g_e^A g_p^V$ vs. $m_{Z'}$ shown in fig. 3b are derived from equation (12) with the use of calculated $E(m_{Z'})$ amplitudes for the Yb $^1S_0 \to {}^3D_1$ transition.

The resulting constraints on $g_e^A g_p^V$ are combined [through equation (10)] with constraints on the effective $g_e^A g_N^V$ coupling, that come from the analysis of the Cs PV results [4]. This analysis involves application of equation (11) for the Cs $Q_W^{new}$=0.65(43) [5] with use of the Cs 6S→7S amplitudes $E(m_{Z'})$. This results in limits on electron-neutron interaction $g_e^A g_n^V$ which are shown in fig. 3c.

The recent measurement of the weak charge of the proton in the Q-weak experiment [32], enables further constraints on $g_e^A g_n^V$ and $g_e^A g_p^V$ in the limit of large $m_{Z'}$ ($m_{Z'}$>1 GeV). The resulting beyond the SM contribution to the proton weak charge from that experiment is $q_p^{new}$=0.0011(45). With $q_p^{new} = Q_W^{new}/Z$, in the large mass-limit $[E(m_{Z'}) \to E_\infty]$, equation (11) yields:

$$g_e^A g_p^V = -q_p^{new} \frac{G_F m_{Z'}^2}{2\sqrt{2}}. \tag{13}$$

Having obtained this constraint, we return to the upper bound on $g_e^A g_N^V$ from the Cs experiment and combine the two results to obtain a large-mass bound on $g_e^A g_n^V$. We list in table II all small-mass and large-mass constraints derived by the analysis presented in this section.

Table II. Upper bounds on electron-proton and electron-neutron interactions mediated by a vector boson Z´of mass $m_{Z'}$. These limits are derived through analysis of the results of different experiments or combinations of these. The large-mass limits $|g_e^A g_p^V|/m_{Z'}^2$ and $|g_e^A g_n^V|/m_{Z'}^2$ are valid for $m_{Z'} > 1$GeV) and the small-mass limits $|g_e^A g_p^V|$ and $|g_e^A g_n^V|$ for ($m_{Z'} < 100$ eV) – see table I in ref. [15].

| | $|g_e^A g_p^V|/m_{Z'}^2$ (GeV)$^{-2}$ | $|g_e^A g_n^V|/m_{Z'}^2$ (GeV)$^{-2}$ | $|g_e^A g_p^V|$ | $|g_e^A g_n^V|$ |
|---|---|---|---|---|
| | Large-mass limit | | Low-mass limit | |
| **Experiment** | | | | |
| Yb PV | $(3.7\pm9)\cdot 10^{-7}$ | ... | $(4.5\pm11)\cdot 10^{-13}$ | ... |
| Yb & Cs PV | ... | $(-2.9\pm6.4)\cdot 10^{-7}$ | ... | $(-3.5\pm7.9)\cdot 10^{-13}$ |
| Q-weak | $(-4.5\pm18.6)\cdot 10^{-9}$ | ... | ... | ... |
| Q-weak & Cs PV | ... | $(-3.1\pm2.6)\cdot 10^{-8}$ | ... | ... |

**Determination of the $\zeta/\beta$ sign** When field imperfections are included in the derivation of the harmonics ratio of equation (7), the resulting ratio (including terms up to 2$^{nd}$ order in small parameters such as these imperfections and $\zeta$) is given by:

$$r_0 = \frac{4}{E_0}\left(E_{dc} + \frac{\zeta}{\beta}\cot\theta + \frac{b_x}{B_Z}e_y\cot\theta - \frac{b_x}{B_Z}e_z + \frac{b_y \zeta}{\beta B_Z} + \frac{b_y \zeta}{\beta B_Z}\cot^2\theta\right), \tag{14}$$

where $e_y$, $e_z$, are the small non-reversing components of the electric field in the interaction region (with the main field applied nominally along x), and $b_x$, $b_y$ are spurious magnetic-field components (the primary component is $B_z$).

We have studied two of the terms contributing to $r_0$, in order to determine the sign of $\zeta/\beta$. Note that the contribution from $E_{dc}$ is sufficient to establish a sign definition for $\zeta/\beta$. We applied a large and positive $E_{dc}$, and adjusted the phase of the lock-in amplifier that measures the part of the 408 nm transition rate oscillating at frequency $\omega$, such that the device output is maximum in magnitude and positive. (The phase of the lock-in measuring the $2\omega$ signal is set so that the corresponding output always has the maximum magnitude and is positive.) This adjustment results in a positive value for $r_0$. Reversing the $E_{dc}$ polarity reverses the sign of $r_0$. With this procedure, the sign of $(\zeta/\beta)\cot(\theta)$ is defined unambiguously.

We have further checked the sign of the 3$^{rd}$ term in equation (14). By enhancing $b_x$ and $e_y$, this term can be made rather large (up to $\times$ 1000 $\zeta/\beta$). We have checked that for $B_z>0$ and enhanced $b_x>0$ and $e_y>0$, $r_0$ is positive when $\theta>0$, and $r_0$ is negative when $\theta<0$. Reversing any of the quantities $b_x$, $e_y$ or $B_z$ causes $r_0$ to reverse sign. This additional test is consistent with the sign definition based on the test with $E_{dc}$ and furthermore, its result is consistent with our definition of the positive sense for the polarization angle $\theta$. Prior to all these tests, and in fact prior to all the PV-experiments, we checked and made sure that all the fields in the experiment (i.e. the voltage polarities of the 10 electrodes of our field plate system, the direction of the all the applied magnetic fields, and the polarization of the 408 nm light) are applied with correct polarity.

In addition to these tests, we made a makeshift measurement of the ratio of M1 transition moment to $\beta$, and compared our result with a prior determination of M1/$\beta$ [33]. This experiment required replacing the field plates shown in fig.2 with others, which were oriented to create an electric field along z (see fig. 2). Since the Stark-M1 interference is suppressed when atoms are excited by a standing-wave field, the PBC mirrors were removed for these measurements. The geometry of applied fields was the same as that employed in [33,34]. Our result for M1/$\beta$ is -23.46(100) V/cm, which agrees in sign with the previous result of -22.3(10) V/cm, thereby offering additional assurance in our PV-effect sign determination. The quoted error in our measurement, of 1 V/cm is due to systematic uncertainty, primarily from the non-optimized construction of the field-plates used.

Finally, to ensure that our determination for the sign of $\zeta/\beta$ is free of possible errors in the computer code used to extract the PV-parameter from raw data or other human error, two of us performed analysis of the same set of data with independently developed codes, and obtained the same results.

To further investigate the sign discrepancy we acquired data with $^{174}$Yb using the method employed in the previous experiment. For this we recorded with the new apparatus 408 nm transition lineshapes like that shown in fig. 4, under all combinations of parity-reversals employed in the 2009 work. One of us used the original data analysis script written for the previous experiment, to perform lineshape fitting in the newly acquired spectra and extract the parameter $\zeta/\beta$, exactly as it was done to obtain the results presented in [10,11]. This analysis yielded a $\zeta/\beta$ value of 21.1(3.6) mV/cm, consistent in magnitude with the more precise determination of −23.89(11) mV/cm that we report in this work for $^{174}$Yb, but whose sign, however, was positive, just as the result reported in 2009. Examination of the data-processing script revealed that a minus sign was inadvertently omitted in the data analysis. Taking this omitted sign into account establishes consistency between the sign results of the experiments reported in [10,11] and our present measurements.

**Data availability.** The data that support the findings of this study are available from the corresponding author upon reasonable request.

**References**

29. Stalnaker, J.E., et al. Dynamic Stark effect and forbidden-transition spectral line shapes Phys. Rev. A **73**, 043416 (2006).

30. Dounas-Frazer, D.R., Tsigutkin, K., Family, A., and Budker, D. Measurement of dynamic Stark polarizabilities by analyzing spectral lineshapes of forbidden transitions. Phys. Rev. A **82**, 062507 (2010).

31. Antypas, D., Fabricant, A., and Budker, D. Lineshape-asymmetry elimination in weak atomic transitions driven by an intense standing wave field, Optics Letters **43**, 2241 (2018).

32. The Jefferson Lab Qweak Collaboration. Precision measurement of the weak charge of the proton. Nature **557**, 207 (2018).

33. Stalnaker, J.E., Budker, D., DeMille, D.P., Friedman, S.J., and Yashchuk, V.V. Measurement of the forbidden 6s$^2$ $^1$S$_0$→5d6s $^3$D$_1$ magnetic-dipole transition amplitude in atomic ytterbium. Phys. Rev. A **66**, 031403 (2002).

34. Budker, D., and Stalnaker, J. E. Magnetoelectric Jones dichroism in atoms. Phys. Rev. Lett. **91**, 263901 (2003).